\RequirePackage{lineno}  

\documentclass[aps,prl,twocolumn,superscriptaddress,showpacs]{revtex4}

\usepackage{graphicx}  % Include figure files
\usepackage{dcolumn}   % Align table columns on decimal point
\usepackage{bm}        % bold math
\usepackage{xspace}

\newcommand{\beq}{\begin{equation}}
\newcommand{\eeq}{\end{equation}}
\newcommand{\bbf}{\begin{bfseries}}
\newcommand{\ebf}{\end{bfseries}}
\newcommand{\beqn}{\begin{eqnarray}}
\newcommand{\eeqn}{\end{eqnarray}}
\newcommand{\bi}{\begin{itemize}}
\newcommand{\ei}{\end{itemize}}
\newcommand{\bd}{\begin{description}}
\newcommand{\ed}{\end{description}}
\newcommand{\ben}{\begin{enumerate}}
\newcommand{\een}{\end{enumerate}}
\newcommand{\bHuge}{\begin{Huge}}
\newcommand{\bhuge}{\begin{huge}}
\newcommand{\bLARGE}{\begin{LARGE}}
\newcommand{\bLarge}{\begin{Large}}
\newcommand{\blarge}{\begin{large}}
\newcommand{\eHuge}{\end{Huge}}
\newcommand{\ehuge}{\end{huge}}
\newcommand{\eLARGE}{\end{LARGE}}
\newcommand{\eLarge}{\end{Large}}
\newcommand{\elarge}{\end{large}}

\newcommand{\ppbar}{\mbox{$p\overline{p}$}}

\newcommand{\ccbar}{\mbox{$c\overline{c}$}}
\newcommand{\ttbar}{\mbox{$t\overline{t}$}}

\newcommand{\met}{\mbox{${E\!\!\!\!/_T}$}}

\newcommand{\pt}{\mbox{$p_T$}}

\newcommand{\vecmet}{\vec{\,/\!\!\!\!E}_{T}}

\newcommand{\hplusm}{\mbox{$H^{\pm}$}}

\newcommand{\ptaupm}{\mbox{$\tau^\pm$}}
\newcommand{\zgamma}{$Z/\gamma^{*}$}

\newcommand{\simgt}{\,\hbox{\lower0.6ex\hbox{$\sim$}\llap{\raise0.6ex\hbox{$>$}}}\,}

\newcommand{\simlt}{\,\hbox{\lower0.6ex\hbox{$\sim$}\llap{\raise0.6ex\hbox{$<$}}}\,}

\begin{document} 

\title{ Search for a Very Light $CP$-Odd Higgs Boson \\ in Top Quark Decays 
from $p\bar{p}$ Collisions at $\sqrt{s}$ = 1.96 TeV}

\date{\today}
\affiliation{Institute of Physics, Academia Sinica, Taipei, Taiwan 11529, Republic of China} 
\affiliation{Argonne National Laboratory, Argonne, Illinois 60439, USA} 
\affiliation{University of Athens, 157 71 Athens, Greece} 
\affiliation{Institut de Fisica d'Altes Energies, Universitat Autonoma de Barcelona, E-08193, Bellaterra (Barcelona), Spain} 
\affiliation{Baylor University, Waco, Texas 76798, USA} 
\affiliation{Istituto Nazionale di Fisica Nucleare Bologna, $^{bb}$University of Bologna, I-40127 Bologna, Italy} 
\affiliation{Brandeis University, Waltham, Massachusetts 02254, USA} 
\affiliation{University of California, Davis, Davis, California 95616, USA} 
\affiliation{University of California, Los Angeles, Los Angeles, California 90024, USA} 
\affiliation{Instituto de Fisica de Cantabria, CSIC-University of Cantabria, 39005 Santander, Spain} 
\affiliation{Carnegie Mellon University, Pittsburgh, Pennsylvania 15213, USA} 
\affiliation{Enrico Fermi Institute, University of Chicago, Chicago, Illinois 60637, USA}
\affiliation{Comenius University, 842 48 Bratislava, Slovakia; Institute of Experimental Physics, 040 01 Kosice, Slovakia} 
\affiliation{Joint Institute for Nuclear Research, RU-141980 Dubna, Russia} 
\affiliation{Duke University, Durham, North Carolina 27708, USA} 
\affiliation{Fermi National Accelerator Laboratory, Batavia, Illinois 60510, USA} 
\affiliation{University of Florida, Gainesville, Florida 32611, USA} 
\affiliation{Laboratori Nazionali di Frascati, Istituto Nazionale di Fisica Nucleare, I-00044 Frascati, Italy} 
\affiliation{University of Geneva, CH-1211 Geneva 4, Switzerland} 
\affiliation{Glasgow University, Glasgow G12 8QQ, United Kingdom} 
\affiliation{Harvard University, Cambridge, Massachusetts 02138, USA} 
\affiliation{Division of High Energy Physics, Department of Physics, University of Helsinki and Helsinki Institute of Physics, FIN-00014, Helsinki, Finland} 
\affiliation{University of Illinois, Urbana, Illinois 61801, USA} 
\affiliation{The Johns Hopkins University, Baltimore, Maryland 21218, USA} 
\affiliation{Institut f\"{u}r Experimentelle Kernphysik, Karlsruhe Institute of Technology, D-76131 Karlsruhe, Germany} 
\affiliation{Center for High Energy Physics: Kyungpook National University, Daegu 702-701, Korea; Seoul National University, Seoul 151-742, Korea; Sungkyunkwan University, Suwon 440-746, Korea; Korea Institute of Science and Technology Information, Daejeon 305-806, Korea; Chonnam National University, Gwangju 500-757, Korea; Chonbuk National University, Jeonju 561-756, Korea} 
\affiliation{Ernest Orlando Lawrence Berkeley National Laboratory, Berkeley, California 94720, USA} 
\affiliation{University of Liverpool, Liverpool L69 7ZE, United Kingdom} 
\affiliation{University College London, London WC1E 6BT, United Kingdom} 
\affiliation{Centro de Investigaciones Energeticas Medioambientales y Tecnologicas, E-28040 Madrid, Spain} 
\affiliation{Massachusetts Institute of Technology, Cambridge, Massachusetts 02139, USA} 
\affiliation{Institute of Particle Physics: McGill University, Montr\'{e}al, Qu\'{e}bec, Canada H3A~2T8; Simon Fraser University, Burnaby, British Columbia, Canada V5A~1S6; University of Toronto, Toronto, Ontario, Canada M5S~1A7; and TRIUMF, Vancouver, British Columbia, Canada V6T~2A3} 
\affiliation{University of Michigan, Ann Arbor, Michigan 48109, USA} 
\affiliation{Michigan State University, East Lansing, Michigan 48824, USA}
\affiliation{Institution for Theoretical and Experimental Physics, ITEP, Moscow 117259, Russia}
\affiliation{University of New Mexico, Albuquerque, New Mexico 87131, USA} 
\affiliation{Northwestern University, Evanston, Illinois 60208, USA} 
\affiliation{The Ohio State University, Columbus, Ohio 43210, USA} 
\affiliation{Okayama University, Okayama 700-8530, Japan} 
\affiliation{Osaka City University, Osaka 588, Japan} 
\affiliation{University of Oxford, Oxford OX1 3RH, United Kingdom} 
\affiliation{Istituto Nazionale di Fisica Nucleare, Sezione di Padova-Trento, $^{cc}$University of Padova, I-35131 Padova, Italy} 
\affiliation{LPNHE, Universite Pierre et Marie Curie/IN2P3-CNRS, UMR7585, Paris, F-75252 France} 
\affiliation{University of Pennsylvania, Philadelphia, Pennsylvania 19104, USA}
\affiliation{Istituto Nazionale di Fisica Nucleare Pisa, $^{dd}$University of Pisa, $^{ee}$University of Siena and $^{ff}$Scuola Normale Superiore, I-56127 Pisa, Italy} 
\affiliation{University of Pittsburgh, Pittsburgh, Pennsylvania 15260, USA} 
\affiliation{Purdue University, West Lafayette, Indiana 47907, USA} 
\affiliation{University of Rochester, Rochester, New York 14627, USA} 
\affiliation{The Rockefeller University, New York, New York 10065, USA} 
\affiliation{Istituto Nazionale di Fisica Nucleare, Sezione di Roma 1, $^{gg}$Sapienza Universit\`{a} di Roma, I-00185 Roma, Italy} 

\affiliation{Rutgers University, Piscataway, New Jersey 08855, USA} 
\affiliation{Texas A\&M University, College Station, Texas 77843, USA} 
\affiliation{Istituto Nazionale di Fisica Nucleare Trieste/Udine, I-34100 Trieste, $^{hh}$University of Trieste/Udine, I-33100 Udine, Italy} 
\affiliation{University of Tsukuba, Tsukuba, Ibaraki 305, Japan} 
\affiliation{Tufts University, Medford, Massachusetts 02155, USA} 
\affiliation{Waseda University, Tokyo 169, Japan} 
\affiliation{Wayne State University, Detroit, Michigan 48201, USA} 
\affiliation{University of Wisconsin, Madison, Wisconsin 53706, USA} 
\affiliation{Yale University, New Haven, Connecticut 06520, USA} 
\author{T.~Aaltonen}
\affiliation{Division of High Energy Physics, Department of Physics, University of Helsinki and Helsinki Institute of Physics, FIN-00014, Helsinki, Finland}
\author{B.~\'{A}lvarez~Gonz\'{a}lez$^v$}
\affiliation{Instituto de Fisica de Cantabria, CSIC-University of Cantabria, 39005 Santander, Spain}
\author{S.~Amerio}
\affiliation{Istituto Nazionale di Fisica Nucleare, Sezione di Padova-Trento, $^{cc}$University of Padova, I-35131 Padova, Italy} 

\author{D.~Amidei}
\affiliation{University of Michigan, Ann Arbor, Michigan 48109, USA}
\author{A.~Anastassov}
\affiliation{Northwestern University, Evanston, Illinois 60208, USA}
\author{A.~Annovi}
\affiliation{Laboratori Nazionali di Frascati, Istituto Nazionale di Fisica Nucleare, I-00044 Frascati, Italy}
\author{J.~Antos}
\affiliation{Comenius University, 842 48 Bratislava, Slovakia; Institute of Experimental Physics, 040 01 Kosice, Slovakia}
\author{G.~Apollinari}
\affiliation{Fermi National Accelerator Laboratory, Batavia, Illinois 60510, USA}
\author{J.A.~Appel}
\affiliation{Fermi National Accelerator Laboratory, Batavia, Illinois 60510, USA}
\author{A.~Apresyan}
\affiliation{Purdue University, West Lafayette, Indiana 47907, USA}
% ...
\collaboration{CDF Collaboration\footnote{With visitors from $^a$University of Massachusetts Amherst, Amherst, Massachusetts 01003,
$^b$Istituto Nazionale di Fisica Nucleare, Sezione di Cagliari, 09042 Monserrato (Cagliari), Italy,
$^c$University of California Irvine, Irvine, CA  92697, 
$^d$University of California Santa Barbara, Santa Barbara, CA 93106
$^e$University of California Santa Cruz, Santa Cruz, CA  95064,
$^f$CERN,CH-1211 Geneva, Switzerland,
$^g$Cornell University, Ithaca, NY  14853, 
$^h$University of Cyprus, Nicosia CY-1678, Cyprus, 
$^i$University College Dublin, Dublin 4, Ireland,
$^j$University of Fukui, Fukui City, Fukui Prefecture, Japan 910-0017,
$^k$Universidad Iberoamericana, Mexico D.F., Mexico,
$^l$Iowa State University, Ames, IA  50011,
$^m$University of Iowa, Iowa City, IA  52242,
$^n$Kinki University, Higashi-Osaka City, Japan 577-8502,
$^o$Kansas State University, Manhattan, KS 66506,
$^p$University of Manchester, Manchester M13 9PL, England,
$^q$Queen Mary, University of London, London, E1 4NS, England,
$^r$Muons, Inc., Batavia, IL 60510,
$^s$Nagasaki Institute of Applied Science, Nagasaki, Japan, 
$^t$National Research Nuclear University, Moscow, Russia,
$^u$University of Notre Dame, Notre Dame, IN 46556,
$^v$Universidad de Oviedo, E-33007 Oviedo, Spain, 
$^w$Texas Tech University, Lubbock, TX  79609, 
$^x$IFIC(CSIC-Universitat de Valencia), 56071 Valencia, Spain,
$^y$Universidad Tecnica Federico Santa Maria, 110v Valparaiso, Chile,
$^z$University of Virginia, Charlottesville, VA  22906,
$^{aa}$Yarmouk University, Irbid 211-63, Jordan,
$^{ii}$On leave from J.~Stefan Institute, Ljubljana, Slovenia, 
}}
\noaffiliation

\begin{abstract}
We present the results of a search for a very light  $CP$-odd
Higgs boson $a_1^0$ originating  
from top quark decays $t \rightarrow H^{\pm}b \rightarrow W^{\pm^{(*)}}a_1^0\,b$, 
and subsequently decaying into~$\tau^+\tau^-$.
Using 
a data sample corresponding to an integrated luminosity
of 2.7~fb$^{-1}$
collected by the CDF~II detector
in $p\bar{p}$ collisions at 1.96~TeV, we perform 
a search for events containing a lepton, three or more jets, and an additional isolated track 
with transverse momentum in the range
3 to 20 GeV$/c$.
Observed events are consistent with background sources,
 and 95\% C.L. limits are set on the branching
ratio of $t \rightarrow \hplusm b$ for various masses of~\hplusm~and~$a_1^0$.

\end{abstract}


\maketitle

%------------------------------------------------------------------
% Introduction ---------------------------------------------------------
%------------------------------------------------------------------

The Higgs boson is the last unobserved particle of the standard model (SM)~\cite{SM_Higgs}.
In the SM the Higgs boson mass is unstable to quantum corrections. This problem is
naturally solved in supersymmetric models~\cite{SUSY}.
In these theories the Higgs boson sector is more complicated. 
The minimal supersymmetric  extension of the standard model (MSSM) 
contains five Higgs bosons: 
a light and a heavy $CP$-even Higgs ($h$ and $H$), 
a $CP$-odd Higgs ($A$), and a pair of charged bosons~($H^\pm$). 
The next-to-minimal supersymmetric model (NMSSM)~\cite{nonSM-Higgs} further extends the MSSM 
to include an additional $CP$-even and $CP$-odd neutral Higgs bosons. 
In the NMSSM the lightest $CP$-odd Higgs $a^0_1$ can 
be below the $b\bar{b}$ threshold,
so that the
$a^0_1$ boson decays only into~$\tau^+\tau^-, \ccbar$,  or $gg$.

The existence of the very light $a^0_1$ boson has
two important implications.
First, the decay mode of the SM-like Higgs $h\to a^0_1a^0_1$ becomes dominant 
and other SM decay 
rates are decreased,
so that 
the SM-like Higgs boson avoids the LEP~II direct limit~\cite{Higgs_limit}.
The light SM-like Higgs helps to solve the naturalness and 
fine-tuning problems arising in the MSSM~\cite{nonSM-Higgs}.
In addition, 
the charged Higgs boson must not be much heavier than the $W$ boson,
which helps to reconcile 
apparent discrepancies in the LEP lepton universality measurements~\cite{radovan}.
Such a charged Higgs boson could appear in top quark decays $t\to H^\pm b$, escaping current 
limits~\cite{chHiggs_limits} due to a new open decay mode $H^\pm  \rightarrow W^{\pm^{(*)}}a^0_1$,
which has not been investigated before.
This motivates a search for 
$a^0_1$ bosons in decays of top quarks.

In the $\ppbar$ collisions at the Fermilab Tevatron the top quarks are produced mainly in pairs, and 
within the SM almost always decay into a $W$ boson and a $b$-quark.
The NMSSM scenario considered above differs from the SM process
by the presence of one or two $a_1^0$~bosons in the final state.  
As the $a^0_1$ boson decay products are expected to have low momenta,
these could remain undetected without affecting the measurements 
of the $\ttbar$ cross section and properties of the top quark. 

In this Letter we report on the first search for a light  
$CP$-odd Higgs boson $a^0_1$ in decays of top quarks through the
intermediate charged Higgs boson $t \rightarrow H^{\pm}b  \rightarrow W^{\pm^{(*)}}a^0_1b$
assuming $a^0_1 \to \tau^+\tau^-$.
We analyze a data set 
corresponding to an integrated luminosity of~2.7~fb$^{-1}$ of 
$\ppbar$ collisions at $\sqrt{s} = 1.96$~TeV
collected by the Collider Detector at Fermilab (CDF~II)~\cite{cdf}, 
searching in candidate $\ttbar$ events for the presence of low-$p_T$
tracks~\cite{cdf-coord}  that could be attributed to $\tau$-decay products.

We select candidate $\ttbar$ events using criteria developed for a $\ttbar$
cross section measurement~\cite{ttbar_xsec}.
The data events used in the analysis are collected
by triggers that identify at least one high-$p_T$ electron or muon candidate
using the online data acquisition system.
Subsequently, each event is required to have a single isolated 
$e$ or $\mu$ with $p_T > 20$ GeV/$c$ and $|\eta| < 2.0(1.0)$ for $e(\mu)$~\cite{lep-id, isol}.
We require missing transverse energy $\met > 20$~GeV~\cite{met_def}, 
as evidence of a neutrino from the $W$-boson 
decay, and at
least three jets with $E_T > 20$~GeV and $|\eta| < 2.0$, 
reconstructed using a fixed cone algorithm of radius
$R \equiv \sqrt{ (\Delta \eta)^2 + (\Delta \phi)^2} = 0.4$~\cite{jet-energy}.
Backgrounds to $\ttbar$ production are reduced 
by requiring at least one 
of the jets to be identified as a $b$-quark candidate using the presence of a displaced secondary vertex~\cite{secvtx},
and by requiring 
the scalar sum  of the transverse energy
of the lepton, $\met$, and jets~($H_T$) to be above 250 GeV.
We observe 1052 events passing these selection criteria, which 
define a pre-signal sample.

The main contribution to the selected sample of events comes from $\ttbar$ production, which 
we model
using the~\textsc{pythia} 6.216 Monte Carlo (MC) generator~\cite{PYTHIA} 
for both SM and new physics top quark decays, assuming $m_t = 172.5$~GeV/$c^2$.
We use an~\textsc{alpgen} 2.13~\cite{ALPGEN} matrix-element generator interfaced to \textsc{pythia} 6.325 
for modeling  W+jets and \zgamma + jets production.
Other sources of events in the pre-signal sample include
diboson production ($WW, WZ, ZZ$)
modeled with~\textsc{pythia} 6.216 MC generator,
and multi-jet QCD events modeled using 
a data-driven approach described in~\cite{QCD}.
The detector response in all MC samples
is modeled by a~\textsc{geant3}-based 
detector simulation~\cite{GEANT}.

We search for $\tau$-leptons from $a^0_1$ boson decays 
in $\ttbar$ candidate events by looking for
at least one low-$p_T$ (3 GeV/$c < p_T < $ 20 GeV/$c$) 
track
in the central detector region $|\eta| < 1.1$.
The track must be well-measured, i.e. it should have
               a sufficient number of hits in the tracking chamber.
               To ensure that the track is consistent with being produced
		in a~$p\bar{p}$  collision, 
		the distance of closest approach of the track with respect 	
		to the beam axis is required to be small.
The track must also originate from the same $\ppbar$ interaction 
as the isolated lepton by requiring that
$|z_{track} - z_{lepton}| < $ 5 cm, 
 where the $z$-coordinate corresponds to 
 the point of the closest approach to the nominal beamline. 
To suppress backgrounds from jets the candidate track is required to be
isolated from other tracks in the event.
We sum the $p_T$ of every well-measured track with $p_T > 0.5$~GeV/$c$,
including the candidate track, within a cone of $\Delta R < 0.4$ around the 
candidate and with a $z$-position of origin within 5 cm of the candidate track $z$.
We require that the ratio of the candidate track $p_T$ 
to the sum $p_T$ of tracks in the cone be at least 0.9.
We also ensure that the track is  
not within $\Delta R < 0.4$ of the 
lepton ($e$ or $\mu$) or a jet, used to define the $t\bar{t}$ candidate.

The isolated tracks can arise from the hard parton-parton interaction producing 
the high-$p_T$ lepton candidate as well as from the "underlying event" (UE).  In what follows, 
we include  in our UE definition contributions from additional simultaneous proton-antiproton collisions.
Non-UE isolated tracks come from physics processes where more
than one lepton is produced but only one is identified, such as 
from \zgamma$\to \ell^+\ell^-$~events where one lepton 
triggers the event, while the other one has a \pt~below 20 GeV/$c$,
or is a  \ptaupm~that leaves a low-\pt~track. We use simulated events to model the track $p_T$ spectra corresponding to 
leptons from the vector boson decays.

We use data to model the characteristics of UE tracks.  
We 
analyze several different 
data samples to verify that
the UE track \pt~spectrum is 
independent of the data source.
We select $Z$ boson candidates by requiring events to have 
two leptons (``dilepton events")
with an invariant mass 
consistent with a $Z$ boson.
We also study ``lepton + jets" events 
by requiring only one lepton candidate, significant missing transverse energy, plus one or two jets.
This data sample is dominated by events 
from $W$ boson plus associated jets production.
We also analyze several data samples of QCD multi-jet events 
collected by triggers that identify at least one jet. 
Each sample requires a different jet $E_T$ threshold.

\begin{figure}[htbp]
\begin{center}
\includegraphics[width=7.5cm]{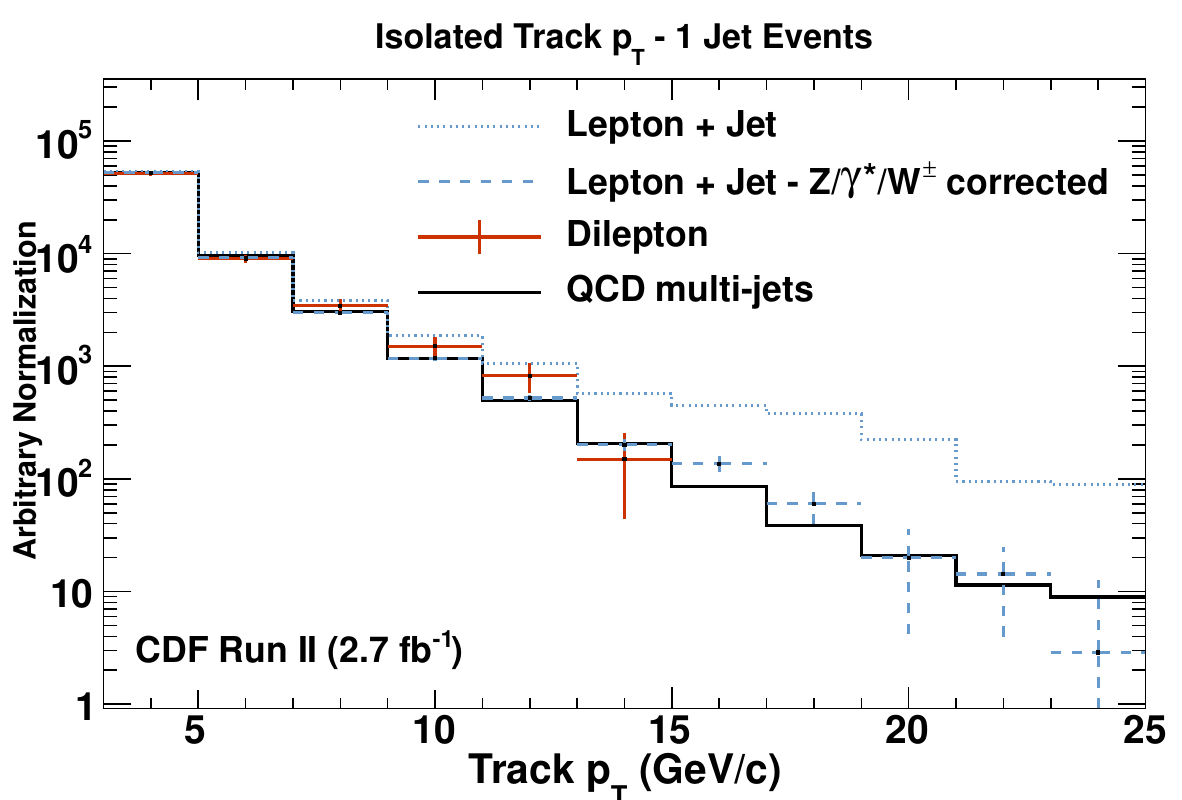}
\caption{ The isolated track \pt~spectra from lepton+jets, dilepton, and QCD multi-jet data 
samples for events with exactly one jet. %, b) with two jets.
%          In both plots the multi-jet sample's track \pt~spectrum is formed from events with two jets.
        }
\label{fig:pretag_track_pt_comp}
\end{center}
\end{figure}

The fraction of events in which
UE tracks satisfy our selection 
criteria is about 7.5\%, and is consistent between samples within 15\% relative uncertainty.
The \pt~spectra of the isolated tracks for different data
           samples normalized to the same area are shown in   
Fig.~\ref{fig:pretag_track_pt_comp}. 
The track \pt~spectrum for lepton+jets events is corrected by subtracting contributions from
tracks corresponding to real leptons from \zgamma , diboson, and $\ttbar$ events.
This is done by accounting for tracks originating from a $W^{\pm}$ or $Z$ boson 
in our MC samples,
where the reconstructed track is traced back to 
the charged particle in the decay 
chain of the vector boson.
In Fig.~\ref{fig:pretag_track_pt_comp} 
both corrected and uncorrected track $p_T$ spectra are shown.
After the correction the \pt~spectra agree with those from dilepton and QCD multi-jet events.

We tested the data to determine whether there are any correlations between the $p_T$ spectra 
of isolated tracks
and other parameters of the event. 
The only correlation we found 
was with the $H_T$ of the event.
We account  for this correlation as described further in the text.
A number of cross-checks that we performed include
comparison of  the UE track \pt~spectra
for different QCD multi-jet samples, as well as a study of dependence on the
jet multiplicity, the number of primary vertices, and presence of the $b$-tag in the events.
We observed no statistically significant difference in the track \pt~spectra in these studies. 

We perform the search for $t \rightarrow H^{\pm}b \rightarrow W^{\pm^{(*)}}a_1^0\,b$ decays 
by fitting the observed isolated track \pt~distribution to the combination of the UE, non-UE SM, 
and the new physics signal track \pt~spectra.
We use the UE track \pt~distribution from QCD multi-jet data events 
to model the UE contribution,
and allow the rate of UE tracks to float freely in the fit.
For the MC-modeled background processes we consider
isolated tracks only from the vector boson decays.
In case an event has more than one track satisfying the isolation and the track quality criteria, we 
select the track with the highest \pt.  
We use the UE track \pt~distribution measured in data
to correct all MC track $p_T$ spectra to account for the probability of the highest-$p_T$ track to come
from the underlying event.

% ======================================================================
%\section{\zgamma Cross Section using Lepton +Jets Data}  \label{sec:lj_zxs_fit}
% ======================================================================

Prior to performing the fit in the signal region, we test our procedure in the 
control region defined by events with one lepton plus one or two jets. In this region the dominant
non-UE contribution is from  \zgamma $\to e^+e^-$ or \zgamma $\to \mu^+\mu^-$ ~events,  where 
the second lepton from \zgamma~is not identified but passes our isolated track requirements,
or \zgamma $\to \tau^+\tau^-$ events where one $\tau$ decays leptonically and is 
identified as an electron or muon, 
and the other one is identified as an isolated track.
The lepton track $p_T$ spectra from 
$Z/\gamma^\star$ 
decays are on average 
more energetic than
the UE track \pt, 
and assuming the UE-only hypothesis
an excess of events is expected  in the tails of the observed isolated track 
\pt~distribution.
We test whether we are able to observe the excess of events attributed to \zgamma $\to \ell^+\ell^-$
events at the rate consistent with the expectation.
The expected number of events from the \zgamma $\to \ell^+\ell^-$ process is obtained using the MC 
normalized to data under the Z mass peak.

\begin{figure}[htbp]
\begin{center}
\includegraphics[width=7.5cm]{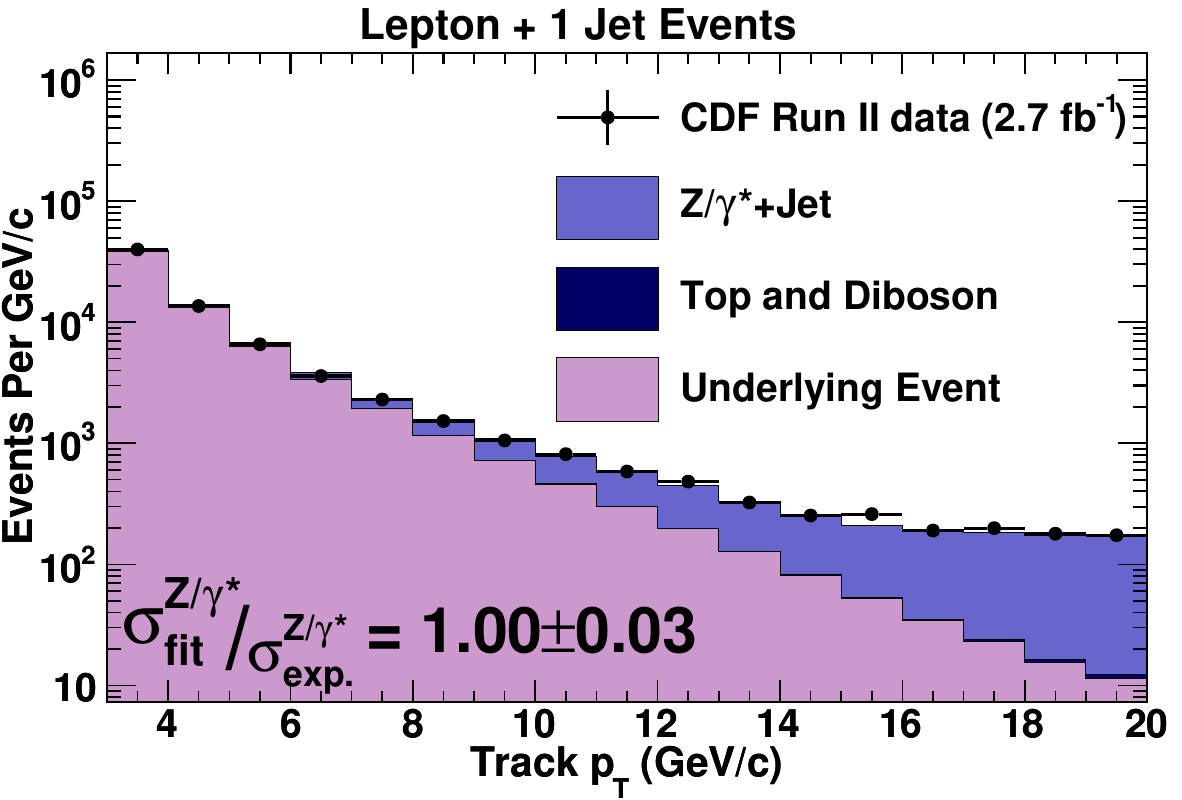}
\includegraphics[width=7.5cm]{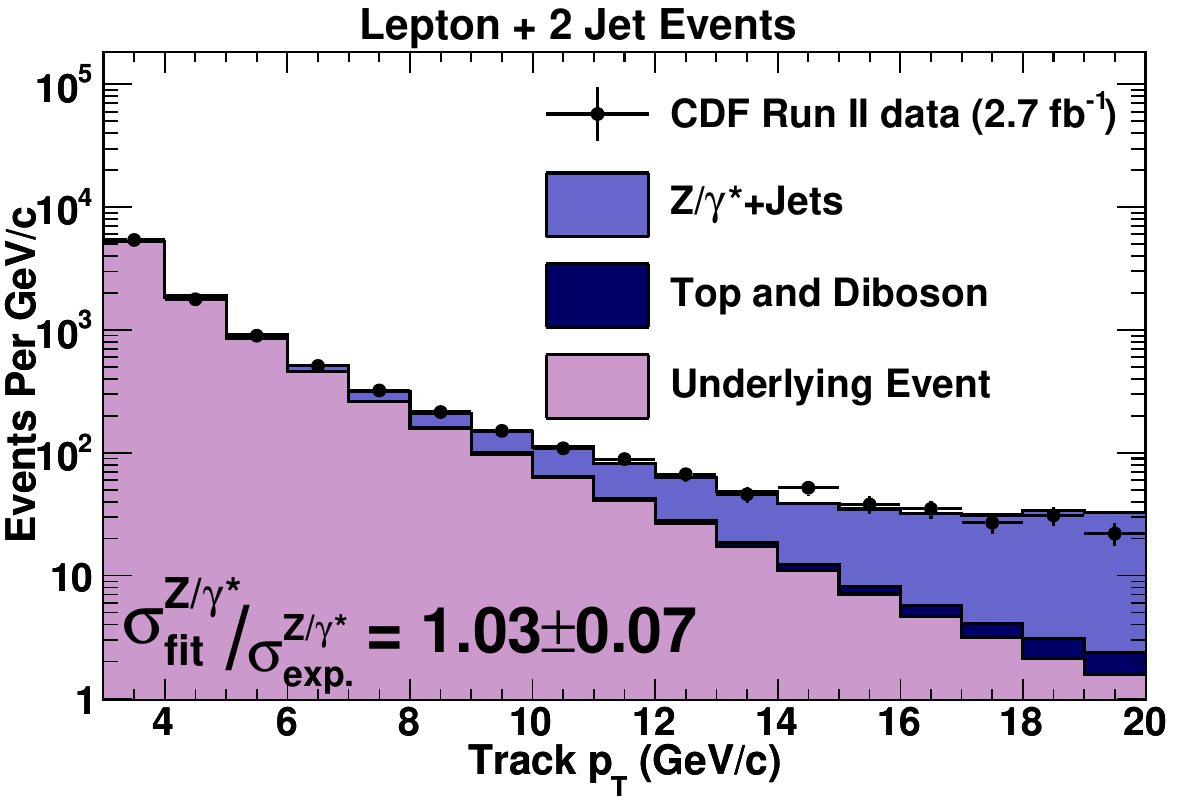}

\caption{ The isolated track $p_T$ spectrum fitted for \zgamma +jets cross section in lepton + jets data events with one jet
		and two jets separately. In both cases the fit results are consistent with expected 
		\zgamma contribution.}
\label{fig:pretagg_zxs_fits}
\end{center}
\end{figure}

We perform 
a log-likelihood fit to the observed isolated track \pt~spectrum, 
with 
UE and \zgamma~rates completely unconstrained, and 
other MC-based contributions (top and dibosons) constrained to 
be within 
their theoretical expectations.
The fit is performed in the range $3 \leq p_T \leq 20 \textrm{ GeV}/c$ separately for 
events with one  
and two jets.
The results of the fit are presented in Fig.~\ref{fig:pretagg_zxs_fits}.
The extracted \zgamma~contribution matches the expectations 
within the statistical uncertainties. 

% ======================================================================
%\section{ Limit Setting Procedure and Systematic Uncertainties} \label{sec:limits}
% ======================================================================
We then proceed to fit in the signal region, and
employ the CL$_{S}$ likelihood ratio test statistic~\cite{cls} to 
quantify the search results.
The systematic uncertainties enter the CL$_{S}$ fit as Gaussian-constrained nuisance parameters.

The $\ttbar$ contribution is obtained from the data using the same technique
as in the $\ttbar$ cross section measurement~\cite{ttbar_xsec}. 
The uncertainty on the expected $\ttbar$ event yield is due to 
the lepton identification and triggering (2\%), $b$-tagging efficiency (5\%), 
the jet energy scale (5\%),
the uncertainty in the estimate of backgrounds to $\ttbar$ (3\%), 
and limited data statistics (6\%) accounting for the total $\ttbar$ normalization uncertainty of 10\%.

The \zgamma + heavy flavor contribution is normalized to data under the $Z$ mass peak, 
with the dominant uncertainty 
due to limited statistics of $Z +$ tagged jet events in data~(8\%).
The uncertainty on the diboson ($VV$) background is due to NLO calculations~\cite{campbell}
and parton distribution functions, taken conservatively to be 10\%,
luminosity~(6\%), and the jet energy scale~(20\%).

Since we require the isolated track not to be within a reconstructed jet, 
the systematic uncertainty in the jet energy scale leads to events migrating 
to/from the signal region, which results in an additional~3\% 
uncertainty for all MC-based backgrounds.
The uncertainty on the isolated track efficiency is 3\%, and is determined using \zgamma events.

The largest variations in the UE isolated track \pt~spectrum come from varying the $H_T$ 
requirement for the candidate sample.
We use the shapes obtained from multi-jet data for very low and very high $H_T$, 
and interpolate these distributions to obtain an intermediate shape. The interpolation is
parametrized with a Gaussian-constrained nuisance parameter and integrated into the fit.
We allow the UE track \pt~distribution to change in the fit according to the value of 
this nuisance parameter~\cite{morph}.

% ==============================================================================
%\section{ Results }
% ==============================================================================

\begin{figure}[htbp]
\begin{center}
\includegraphics[width=7.5cm]{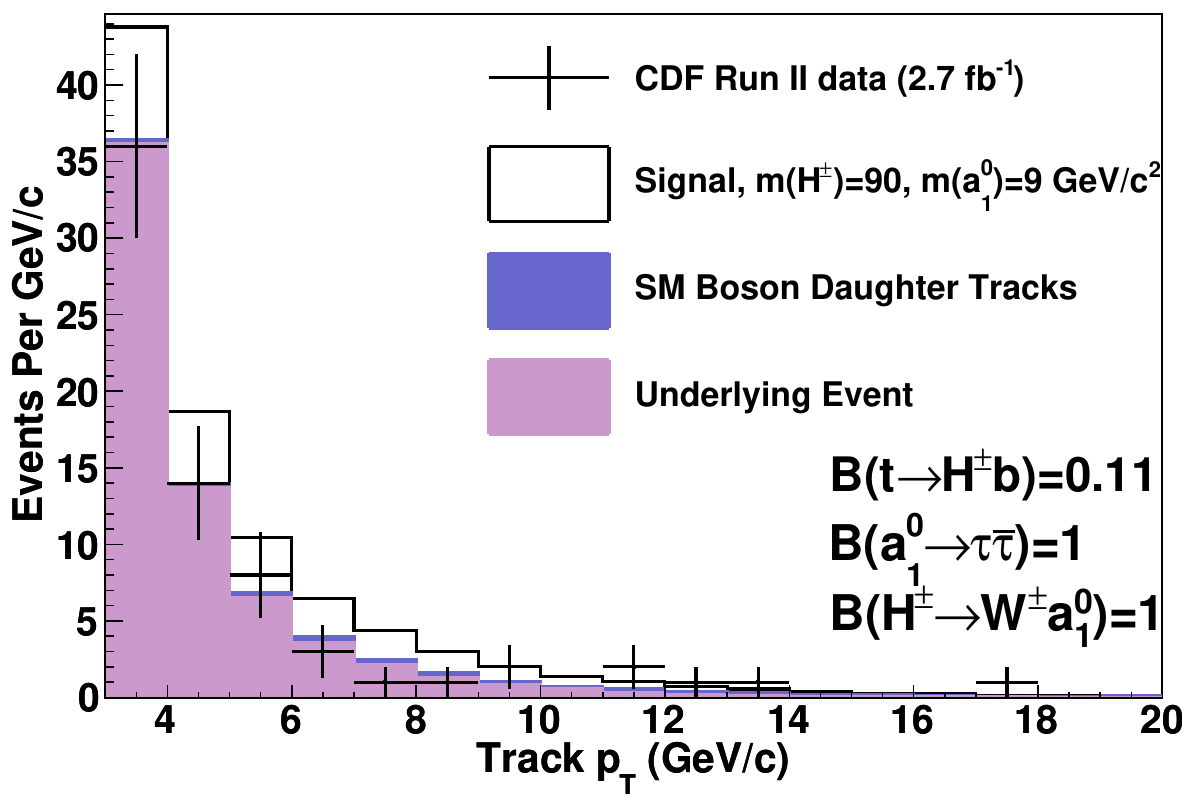}
\caption{ The isolated track \pt~spectrum. The contribution 
from non-SM top decays corresponds to an example scenario
that is excluded at 95\% C.L.
	}
\label{fig:fit_signal}
\end{center}
\end{figure}

The expected event yields in the signal region are presented in Table~\ref{tab:result}. 
The first row in the table represents the numbers of expected and observed events 
before the isolated track requirement. 
The second row shows the event yields after the isolated track requirement, 
where events are
categorized based on the origin of the isolated track.
The quoted event yield due to the UE corresponds to 
the expected rate, while the actual normalization is obtained from the fit to the isolated 
track \pt~spectrum in data, as can be seen in Fig.~\ref{fig:fit_signal}.

\begin{figure}[htbp]
\begin{center}
\includegraphics[width=7.5cm]{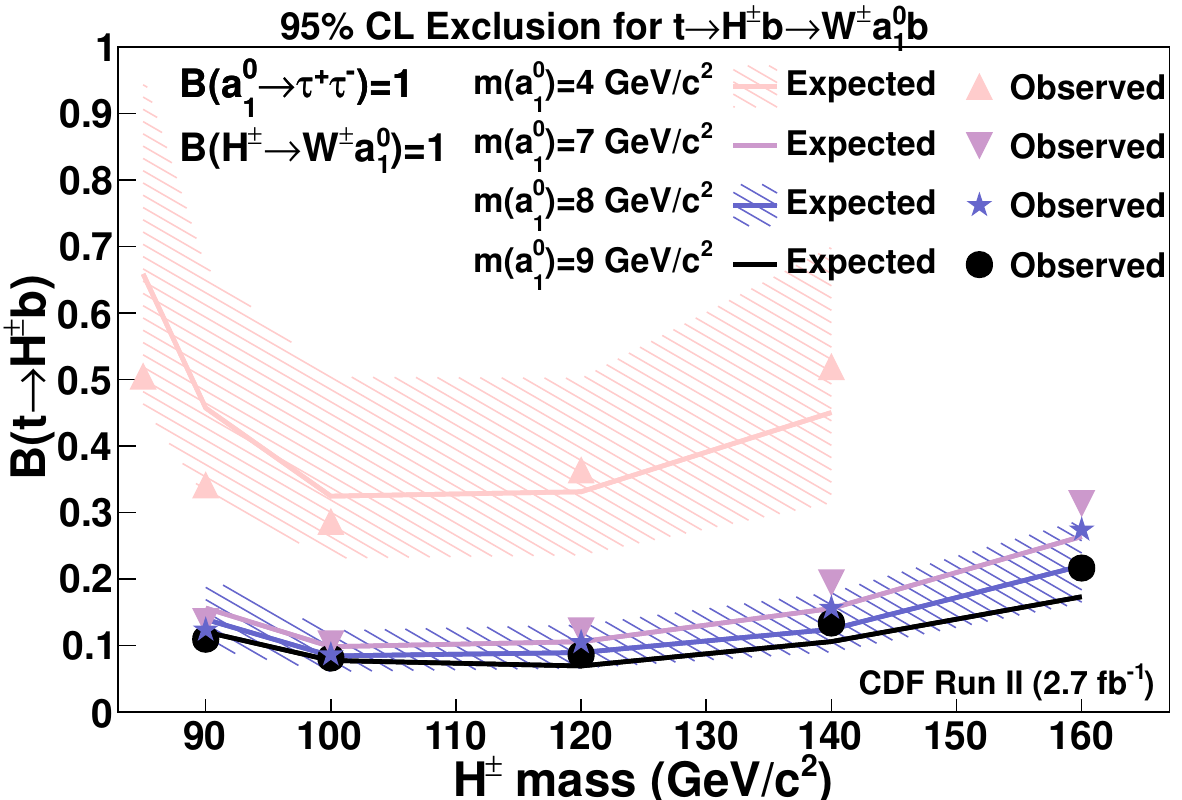}
\caption{
	Observed and expected limits (with $\pm 1$ $\sigma$ error band) on the branching ratio of $t \to H^+ b$.  }
\label{fig:exp_exclusion}
\end{center}
\end{figure}

Figure~\ref{fig:fit_signal} shows that the data are well described by SM background sources. We set 95\% confidence level (C.L.) upper limits on
the branching ratio of $t \to H^+ b$  under the assumption that the branching ratios ${\cal B} (a^0_1 \to \tau^+\tau^-) = 100\%$
and ${\cal B} (H^\pm \to W^\pm a^0_1) = 100\%$. The expected and observed 95\% C.L. limits as a function of 
$m_{H^\pm}$ and $m_{a_1^0}$
are shown in Fig.~\ref{fig:exp_exclusion}. For a given mass of $CP$-odd Higgs boson $a_1^0$ we exclude 
the branching ratios of ${\cal B}(t \to H^+ b)$ above the respective curve shown in the plot.
For an $a_1^0$  boson with mass of 9 GeV/$c^2$, we exclude a ${\cal B}(t \to H^+ b)> 0.20$ at 95\% C.L. for $H^+$ masses between 90 and 160 GeV/$c^2$.
These are the first limits on the branching ratio of $t \to H^+ b$ in this decay mode.

In conclusion, we have presented a search for non-SM top decays 
$t \rightarrow H^{\pm}b \rightarrow W^{\pm^{(*)}}a^0_1\,b$
within the NMSSM scenario using a data sample corresponding
to 2.7 fb$^{-1}$ of integrated luminosity in 1.96~TeV
$\ppbar$ collisions.
We see no evidence of $\tau$'s from light Higgs $a^0_1$ decays, 
and set the world's first limits on the branching ratio of $t \to H^+ b$ 
in this mode, assuming
 ${\cal B} (a^0_1 \to \tau^+\tau^-) = 100\%$
and ${\cal B} (H^\pm \to W^\pm a^0_1) = 100\%$. 

\begin{table}
\begin{tabular}{cccccccc}
\multicolumn{7}{c}{Events per 2.7 fb$^{-1}$}    \\ \hline
       \hline  
                                          $\ttbar$ &  QCD,$W$ & $VV$ & $Z/\gamma^*$     &    UE  & Total & Data      \\ \hline                        
       805   &   215 & 11 & 19  & - & 1049 & 1052   \\
   2.6$\pm$0.3 & - & 0.1$\pm$0.0 & 0.7$\pm$0.1 & 79$\pm$12 & 83$\pm$12 & 70 \\

    \hline\hline
       \end{tabular}  
 \caption{ Expected event yields in 2.7 fb$^{-1}$ before the track requirement (first row), 
          and with at least one isolated track
          (second row) with 3 GeV/$c  < p_{T} < 20\,\textrm{GeV}/c$. 
          In the second row events are categorized based on the origin of the isolated track.
	  The number of UE events is the expected number of events before the fit to the track $p_T$ spectrum.
           }
  \label{tab:result}
\end{table}

% Acknowledgements-------------------------------------------------------
We thank the Fermilab staff and the technical staffs of the participating institutions for their vital contributions. This work was supported by the U.S. Department of Energy and National Science Foundation; the Italian Istituto Nazionale di Fisica Nucleare; the Ministry of Education, Culture, Sports, Science and Technology of Japan; the Natural Sciences and Engineering Research Council of Canada; the National Science Council of the Republic of China; the Swiss National Science Foundation; the A.P. Sloan Foundation; the Bundesministerium f\"ur Bildung und Forschung, Germany; the Korean World Class University Program, the National Research Foundation of Korea; the Science and Technology Facilities Council and the Royal Society, UK; the Institut National de Physique Nucleaire et Physique des Particules/CNRS; the Russian Foundation for Basic Research; the Ministerio de Ciencia e Innovaci\'{o}n, and Programa Consolider-Ingenio 2010, Spain; the Slovak R\&D Agency; the Academy of Finland; and the Australian Research Council (ARC).

% Bibliography ----------------------------------------------------------


\begin{thebibliography}{99}
\bibitem{cdf-coord} CDF uses a cylindrical coordinate system with the $z$ axis along the proton beam axis. 
 				$\theta$ is the polar angle relative to the proton beam direction,
				and 	$\phi$ is the azimuthal angle.
				Pseudorapidity is  defined as $\eta \equiv -\ln(\tan\frac{\theta}{2})$, 
				 while transverse momenta and energies of particles are defined as 
                                      $p_T = |p|\sin\theta$ and $E_T = E\sin\theta$, respectively.
\bibitem{isol} The lepton is defined as isolated if the energy deposit within 
$R=0.4$ cone of the lepton momentum $p_T$ is less than 10\% of the lepton $p_T$.
\bibitem{met_def} The missing transverse energy is defined as $\met = |\vecmet| = | - \sum_i E_T^i \vec{n_i}|$ 
where $\vec{n}_i$ is a unit vector in the plane transverse to the beam direction pointing
from the event vertex to the $i$-th calorimeter tower. The missing transverse energy is corrected 
for the escaping muon momenta when muon candidates are present.
\end{thebibliography}
\end{document}